\documentclass[
aps,
pra,
twocolumn,
groupedaddress,
raggedfooter
]{revtex4-2}

\usepackage{amsmath}
\usepackage{amsfonts}
\usepackage{graphicx}
\usepackage{amssymb}
\usepackage{textcomp}
\usepackage{bm}
\usepackage{color}
\usepackage{dsfont}

\usepackage{pgfplots}

\usepackage{siunitx}

\usepackage{mathtools}

\newcommand{\fig}[1]{Fig.\,\ref{#1}}

\newcommand{\eqP}[1]{(\ref{#1})}

\newcommand{\bitem}{\begin{itemize}}
\newcommand{\eitem}{\end{itemize}}

\newcommand{\bti}{\begin{tikzpicture}}
\newcommand{\eti}{\end{tikzpicture}}

\newcommand{\ket}[1]{\left| #1 \right>} 
\newcommand{\bra}[1]{\left< #1 \right|} 

\newcommand{\Er}{E_{\text{r}}}

\newcommand{\hc}{\text{hc}}
\newcommand{\I}{\text{i}}

\newcommand{\E}{\text{e}}

\newcommand{\Veff}{V_{\text{eff}}}

\newcommand{\Mod}[1]{\ (\mathrm{mod}\ #1)}

\newlength{\OneColumnPRLWidth}
\DeclareSIUnit\gauss{G}

\begin{document}

 \title{
Simulating two-dimensional dynamics
within a large-size atomic spin
 }

 \author{Aurélien Fabre}
 \author{Jean-Baptiste Bouhiron}
 \author{Tanish Satoor}
\author{Raphael Lopes}
\author{Sylvain Nascimbene}
\email{sylvain.nascimbene@lkb.ens.fr}
 \affiliation{Laboratoire Kastler Brossel,  Coll\`ege de France, CNRS, ENS-PSL University, Sorbonne Universit\'e, 11 Place Marcelin Berthelot, 75005 Paris, France}
 \date{\today}

  \begin{abstract}
Encoding a dimension in the internal degree of freedom of an atom provides an interesting tool for quantum simulation, facilitating the realization of artificial gauge fields. We propose an extension of the synthetic dimension toolbox, making it possible to encode two dimensions within a large atomic spin.  The protocol combines first- and second-order spin couplings, such that the spin projection $m$ and the remainder $r=m$\;(mod 3) of its Euclidian division by 3  act as orthogonal coordinates on a synthetic cylinder. It is suited for an implementation with lanthanide atoms, which feature a large electronic spin and narrow optical transitions for applying the required spin couplings. This method is useful for simulating geometries with periodic boundary conditions, and engineering various types of topological systems evolving in high dimensions.
 \end{abstract}
 
 \maketitle
 
Ultracold atomic gases provide a versatile playground for the study  of various types quantum many-body physics. The simulation of artificial gauge fields enables the realization of systems exhibiting a non-trivial topological character \cite{dalibard_colloquium_2011,cooper_topological_2019}. A well-developed protocol for their implementation is based on light-induced couplings between the atom motion and its spin. This technique enables the realization of a synthetic dimension, fully encoded in the internal degree of freedom of the atom, namely its electronic and/or nuclear spin \cite{boada_quantum_2012}. The dynamics of atoms subjected to such a spin-orbit coupling can be described by an effective gauge field \cite{celi_synthetic_2014,ozawa_topological_2019}, which has been used to engineer  two-dimensional quantum Hall systems, with one spatial dimension and another synthetic one \cite{mancini_observation_2015,stuhl_visualizing_2015}. Synthetic dimensions are also promising for the realization of high-dimensional systems that would feature a topological character with  no equivalent in lower dimensions \cite{qi_topological_2008,price_four-dimensional_2015,lian_five-dimensional_2016}. 

The most natural implementation of a synthetic dimension consists in  considering the spin projection $m$ of the atomic spin $J$ (with $|m|\leq J$, $m$ integer \cite{Note2}) as the coordinate of an artificial dimension \cite{mancini_observation_2015,stuhl_visualizing_2015}. Motion along this dimension then occurs via spin transitions $m\rightarrow m'$, for example induced by radio-frequency or two-photon optical transitions. The range $|m-m'|$ of spin transitions is then limited by selection rules to nearest ($|m-m'|=1$) or next-nearest ($|m-m'|=2$) neighbor hoppings. This constraint restricts the simulation of periodic boundary conditions to small-spin systems \cite{li2018bose,han_band_2019,liang_coherence_2021}. Indeed, a coupling between stretched states $m=\pm J$ requires $2J$-photon optical transition, which is experimentally unrealistic for $J\gg1$. In the absence of such coupling, the synthetic dimension features sharp edges \cite{mancini_observation_2015,stuhl_visualizing_2015}, such that the bulk physics is limited to  projection states $m$ far enough from edges \cite{chalopin_probing_2020}. 
 The concept of synthetic dimension was also generalized to atomic momentum states \cite{an_direct_2017}, and has also been developed in photonic systems \cite{yuan_synthetic_2018}. Recently, a pair of synthetic dimensions was simulated in a temporally modulated ring resonator  \cite{dutt_single_2020}.
 
\begin{figure}[!t]
\includegraphics[
 trim={2mm 2mm 0 0.cm},
 scale=1
]{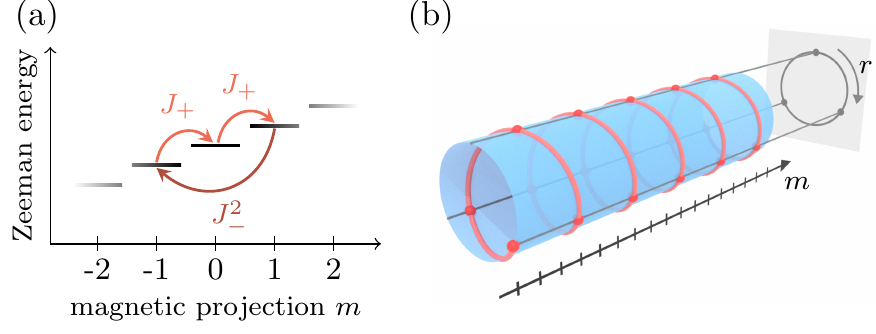}
\caption{
(a) Scheme of the spin transitions between the projection states $\ket{m}$ of an angular momentum $J$. Combining first- and second-order couplings leads to non-trivial 3-cycles $m\rightarrow m+1\rightarrow m+2\rightarrow m$. (b) Scheme of the emergent synthetic cylinder for $J=8$, where the projection $m$ plays the role of the axial coordinate, and the  remainder  $r\equiv m\Mod{3}$ of its Euclidian division by 3 acts as the azimuthal cyclic coordinate.
\label{fig_cylinder}}
\end{figure}
 
In this article, we propose a new protocol to simulate dynamics in two dimensions within the  atomic spin only.  It applies to atomic species possessing a large  spin $J\gg1$. We propose to combine spin couplings of ranks 1 and 2, such that the spin projection $m$ and the remainder $r\equiv m\Mod{3}$ of its Euclidian division by 3 evolve independently, thus acting as the two orthogonal coordinates describing the surface of a cylinder (see \fig{fig_cylinder}). We discuss the conditions of applicability of this description, and the requirements for its practical implementation in cold atom experiments. We also describe its extension for the simulation of quantum Hall physics on a cylinder, with one spatial dimension $x$ and another one encoded in the remainder $r$, which naturally features periodic boundary conditions (the coordinate $m$ adding another degree of freedom, non relevant in this case since it is uncoupled to the $x$ and $r$ dynamics).

\section{Basic description of the protocol\label{section_protocol}}

The protocol combines linear and quadratic spin couplings, described by the Hamiltonian
\begin{align}
H=-U_a\frac{J_+}{J}-U_b\frac{J_-^2}{J(J-1/2)}+\hc\label{eq_H}.
\end{align}
The transitions between magnetic sub-levels $\ket{m}$ induced by these couplings are shown in \fig{fig_cylinder}(a). They enable non-trivial cycles between triples of spin states $m\rightarrow m+1\rightarrow m+2\rightarrow m$, leading to the emergence of a cyclic synthetic coordinate, independent from the magnetic projection $m$, and encoded in the division remainder $r=m\Mod{3}$.

The projection $m$ and remainder $r$ obviously do not evolve independently under the action of either the linear or quadratic spin couplings considered independently. Indeed, the linear coupling $J_+$ increases both $m$ and $r$ by one unit, while the quadratic one $J_-^2$  decreases $m$ by 2 and increases $r$ by one unit. The occurrence of decoupled $m$ and $r$ dynamics relies on the proper combination of both processes. 

In order to understand the condition for independent dynamics, we first give a hand-waving argument
-- a more rigorous treatment being given in section \ref{section_spectrum}. We treat $m$ and $r$ as continuous variables, and approximate the action of the spin operators as 
\begin{align}
\frac{J_++J_-}{J}\psi(m,r)&\simeq \psi(m\!+\!1,r\!+\!1)+\psi(m\!-\!1,r\!-\!1)\\
&\simeq \left(2+\partial_m^2+2\partial_m\partial_r+\partial_r^2\right)\psi(m,r)
\end{align}
and 
\begin{align}
\frac{J_+^2+J_-^2}{J(J\!-\!\tfrac{1}{2})}\psi(m,r)&\simeq \psi(m\!+\!2,r\!-\!1)+\psi(m\!-\!2,r\!+\!1)\\
&\simeq \left(2\!+\!4\partial_m^2\!-\!4\partial_m\partial_r\!+\!\partial_r^2\right)\!\psi(m,r),
\end{align}
at the first non-trivial order in $m$ and $r$.
The Hamiltonian then takes the expression
\begin{align}
 H=&-2(U_a+U_b)-(U_a+4U_b)\partial_m^2\nonumber\\
 &-(U_a+U_b)\partial_r^2-2(U_a-2U_b)\partial_m\partial_r.
\end{align}
The coupling between the $m$ and $r$ dynamics stems from the last term $\propto \partial_m\partial_r$, which cancels for the coupling ratio
\begin{equation}
 U_b/U_a=1/2.\label{eq_ratio}
\end{equation}
Under this condition, the $m$ and $r$ dynamics become approximately separable, mimicking the  motion of a particle on a cylindrical surface with an axial coordinate $m$ and an azimuthal coordinate $r$ (see \fig{fig_cylinder}(b)). Unless explicitly specified, we assume in the following this condition to be fulfilled, and define a single coupling amplitude $U\equiv U_a=2\,U_b$.

\section{Semi-classical analysis and emergence of a synthetic cylinder\label{section_spectrum}}

A more precise understanding of the spin dynamics can be obtained by performing a semi-classical analysis, which is legitimate for a large spin size $J\gg1$. 

\begin{figure}[!t]
\includegraphics[
 trim={2mm 2mm 0 0.cm},
 scale=0.98
]{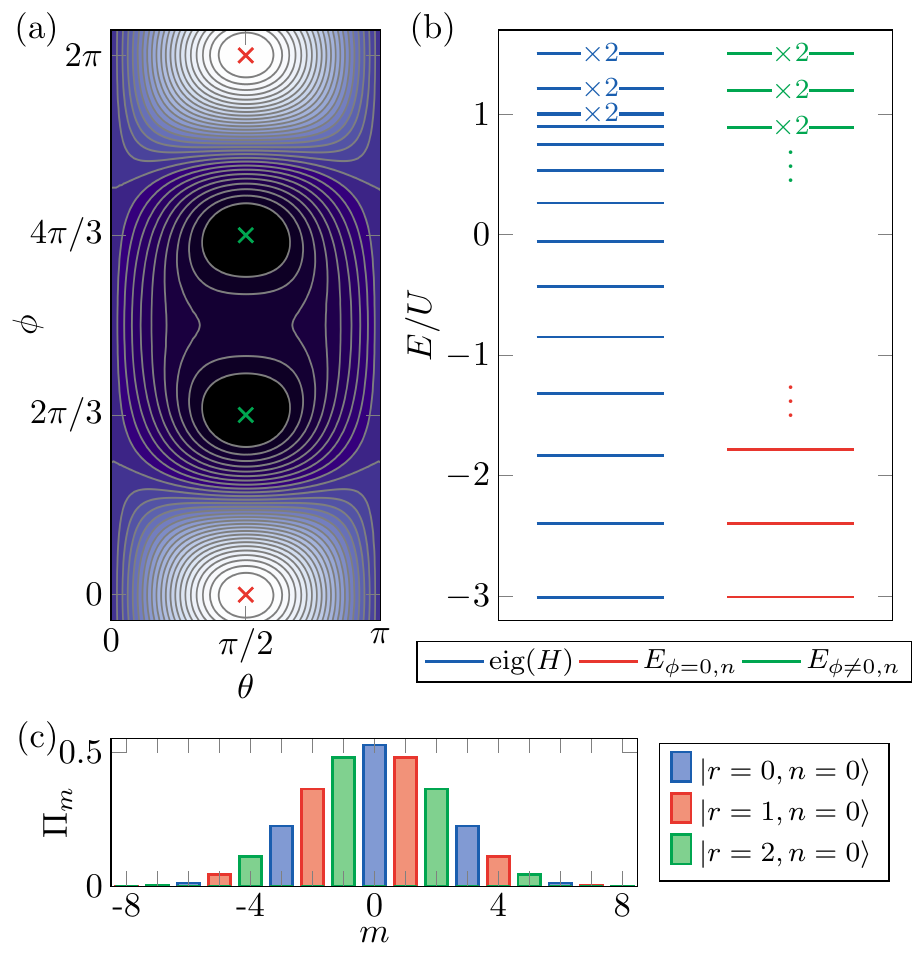}
\caption{
(a) Semi-classical energy functional corresponding to the energy of a coherent spin state of spherical angles $(\theta,\phi)$. The red cross indicates the energy minimum at $\theta=\pi/2$ and $\phi=0$. The green crosses show the two degenerate maxima at $\phi=2\pi/3$ and $4\pi/3$.  (b) Energy levels spectra of the actual Hamiltonian \eqP{eq_H} (blue lines) and of its harmonic approximation around the energy minimum (red lines) or maxima (green lines), for a spin length $J=8$. The label $\times$2 indicates doubly degenerate levels. The degeneracy is not exact for the spectrum of $H$, and the label is shown when two levels are separated by less than the line width. (c) Spin projection probabilities $\Pi_m$ for the states $\ket{r,n}$, with $n=0$  and $r=0,1,2$ (blue, red and green bars).
\label{fig_spectrum}}
\end{figure}

\subsection{Semi-classical ground state}

We first carry out a variational study of the ground state, restricted to the family of coherent spin states. A coherent spin state is defined as a maximally polarized state $\ket{\theta,\phi}$, parametrized by the orientation of its polarization,   labeled by the spherical angles $(\theta,\phi)$ \cite{arecchi_atomic_1972}. The energy associated with a coherent state is described by the functional 
\begin{align}
 E(\theta,\phi)&\equiv\bra{\theta,\phi}H\ket{\theta,\phi}\\
 &=-2U\,\sin\theta\cos\phi-U\,\sin^2\theta\cos(2\phi),\label{eq_functional}
\end{align}
shown in \fig{fig_spectrum}(a). 
It features a single minimum oriented along $x$, that is $\theta=\pi/2,\phi=0$.

\subsection{Harmonic low-energy dynamics}

In order to understand the low-energy dynamics, we expand the Hamiltonian around the semi-classical ground state, assuming that the spin states remain highly polarized along $x$. The $z$ and $y$ spin components then exhibit a commutator
\begin{equation}
[J_z,J_y]=-\I\hbar J_x\simeq -\I\hbar J,
\end{equation}
such that $J_z$ and $(-J_y/J)$ can be considered as canonically conjugated. Expanding the Hamiltonian in powers in these operators, we obtain at the lowest non-trivial order a quadratic Hamiltonian 
\[
H\simeq U\left(-3-\frac{5}{2J}+\frac{4 J_z^2+6J_y^2}{2J^2}\right).
\]
It describes the dynamics of a harmonic oscillator, of spectrum
\begin{align}
 E_{\phi=0,n}&=E_0+n\hbar\omega ,\label{eq_harmonic}\\
 E_0&= [-3+(2\sqrt6-5)/(2J)]U,\\
 \hbar\omega&=(2\sqrt6/J) U,\label{eq_omega}
\end{align}
where $n\geq0$ is an integer and $\omega$ is the effective oscillator frequency. We discuss in Appendix an alternative derivation, based on a Holstein-Primakoff transform of spin operators in terms of a bosonic degree of freedom \cite{holstein_field_1940}.

\subsection{Extension to high energy states}

The variational analysis can also be used to get the highest energy states. The energy functional $E(\theta,\phi)$ exhibits two degenerate maxima, at $\theta=\pi/2$ and $\phi=2\pi/3$ or $4\pi/3$ (see \fig{fig_spectrum}(a)). The dynamics around these  maxima can also be approximated by a harmonic spectrum, which turns out to be linked to the spectrum calculated around the ground state $\phi=0$, as $E_{\phi\neq0,n}=-E_{\phi=0,n}/2$. Overall, the harmonic spectra calculated around the energy minimum ($\phi=0$) or maxima ($\phi=2\pi/3,4\pi/3$) can be recast into a single expression
\begin{equation}\label{eq_spectrum_phi_n}
 E_{\phi,n}=(E_0+n\hbar\omega)\cos\phi, \quad n\in \mathbb{N}, \quad\phi\in\Big\{0,\frac{2\pi}{3},\frac{4\pi}{3}\Big\}.
\end{equation}
We show in \fig{fig_spectrum}(b) a comparison between the spectrum of the actual Hamiltonian \eqP{eq_H} and the approximated spectrum \eqP{eq_spectrum_phi_n}, calculated for $J=8$. The harmonic spectrum accounts well for the first levels above the ground state and the states below the highest energy levels. We checked that the number of levels well described by the harmonic spectrum increases when increasing the spin length $J$, as expected for a semi-classical analysis.

\subsection{Interpretation as a cylindrical geometry}
The spectrum \eqP{eq_spectrum_phi_n} obtained from the semi-classical analysis is relevant to describe spin dynamics at low and high energies, but does not apply in the intermediate energy regime. Still, we consider here the effective spin dynamics restricted to the semi-classical spectrum, and interpret it in terms of motion on  a synthetic cylinder. This approach will become fully justified when coupling the spin to a spatial degree of freedom, such that the three coherent states indexed by $\phi$ occur at low energy on equal footings (see section \ref{section_xr_cylinder}).

The semi-classical spectrum \eqP{eq_spectrum_phi_n}, proportional to $\cos\phi$ with $\phi=0,2\pi/3,4\pi/3$, is reminiscent of the dispersion relation $E(q)\sim-2t\cos(qa)$ of a particle evolving on a one-dimensional ring lattice of length $L$, where $t$ is the tunnel coupling and $a$ is the lattice constant. The quasi-momentum $q$ takes the discrete values $(2\pi j)/L$, with $0\leq j<L/a$ an integer. By analogy, the 3 discrete angles $\phi$ involved in our problem play the role of the momenta conjugated to a cyclic dimension of length $L/a=3$. 

This motivates the definition of a basis of position states $\ket{r,n}$, where $r$ is the coordinate of the synthetic dimension, by the inverse Fourier transform
\begin{equation}
 \ket{r,n}=\frac{1}{\sqrt 3}\sum_{\phi=0,\tfrac{2\pi}{3},\tfrac{4\pi}{3}}\E^{-\I\phi r}\ket{\phi,n}.
\end{equation}
The spin projection probabilities $\Pi_m$ of the states $\ket{r,n}$, shown in \fig{fig_spectrum}(c) for $n=0$, only involve projections $m$ such that $m\Mod{3}=r$, justifying the $r$ notation.
The spectrum \eqP{eq_spectrum_phi_n}, associated to an effective Hamiltonian diagonal in the $\ket{\phi,n}$ basis, can be recast in terms of the $\ket{r,n}$ states as
\begin{align}
 H_{\text{eff}}&=\sum_{n\geq 0}\sum_{\phi=0,\tfrac{2\pi}{3},\tfrac{4\pi}{3}} (E_0+n\hbar\omega)\cos\phi\ket{\phi,n}\bra{\phi,n}\\
 &=\sum_{n\geq 0}\sum_{r=0}^2\frac{E_0+n\hbar\omega }{2}\ket{r+1,n}\bra{r,n}+\hc.
\end{align}
We recognize the Hamiltonian of a particle on a cylinder, with free dynamics along the azimuthal direction $r$, and harmonic trapping along the axis $m$.

\section{Low-energy dynamics}

 \subsection{Excitation protocol}

We illustrate the independent motion along the two directions $m$ and $r$ with simulations of spin dynamics. Starting in the ground state of the Hamiltonian \eqP{eq_H}, we apply a weak perturbation that induces a non-zero velocity either along $m$ or along $r$. The velocity along $m$ is defined as
\begin{align}
 v_m&\equiv\frac{\I}{\hbar}[H,J_z]\\
 &=U_a\frac{\I J_+}{J}+U_b\frac{-2\I J_-^2}{J(J-1/2)}+\hc.
\end{align}
The cyclic coordinate $r$, which can be viewed as an angular variable, cannot be expressed in terms of  an Hermitian operator \cite{barnett_quantum_1990,lynch_quantum_1995}. To obtain the expression of the velocity along $r$, we replace the prefactor $-2$ in front of the $J_-^2$ coupling by 1, to account for the different hopping values $\Delta m=-2$ and $\Delta r=1$. Since the $J_+$ coupling induces identical hoppings $\Delta m=\Delta r=1$, its prefactor remains the same for the two velocities. This leads to the expression
\footnote{
The velocity along $r$ can be recovered using the unitary angle operator $\exp(\I \tfrac{2\pi}{3} J_z)$, from its commutator with the Hamiltonian, as
\begin{align*}
 v_r&\equiv\frac{\I}{\hbar}\frac{1}{\sqrt 3}\left\{\exp\left(-\I \frac{2\pi}{3} J_z\right),\left[H,\exp\left(\I \frac{2\pi}{3} J_z\right)\right]\right\}\\
 &=U_a\frac{\I J_+}{J}+U_b\frac{\I J_-^2}{J(J-1/2)}+\hc,
\end{align*}
which coincides with the expression given in \eqP{eq_vr}.}
\begin{align}
 v_r=U_a\frac{\I J_+}{J}+U_b\frac{\I J_-^2}{J(J-1/2)}+\hc.\label{eq_vr}
\end{align}

The velocity kick along $m$ is applied by evolving a Zeeman field along $z$
\begin{equation}
V_{\text{pert}}^{(m)}=V_z\,J_z,
\end{equation}
corresponding to a linear potential in $m$.

To induce a velocity along $r$, we need to couple the ground state $\ket{\phi=0,n=0}$ to the states $\ket{\phi\neq0,n=0}$. Since the states $\ket{\phi,n=0}$ are coherent spin states spread along the equator with azimuthal angles $2\pi/3$, two states with different angles $\phi$ are very distant in phase space for $J\gg1$, and thus cannot be coupled with low-order spin couplings. To excite the $r$-velocity, we apply a time-dependent perturbation involving the high-order coupling $V_{\text{pert}}^{(r)}(t)=V_r\,\cos(2\pi J_z/3-\alpha t)$. This coupling, diagonal in the $\ket{m}$ projection state basis, is 3-periodic in $m$, such that it takes a value depending on $r$ only, as 
\begin{equation}
V_{\text{pert}}^{(r)}(t)=V_r\,\cos(2\pi r/3-\alpha t).
\end{equation}
This potential corresponds to a perturbation in $r$ moving at the speed $3\alpha/(2\pi)$, which drives the system to a non-zero velocity $\langle v_r\rangle\neq 0$.

 \begin{figure}[!t]
 \includegraphics[
  trim={3mm 2mm 0 0.cm},scale=0.87
 ]{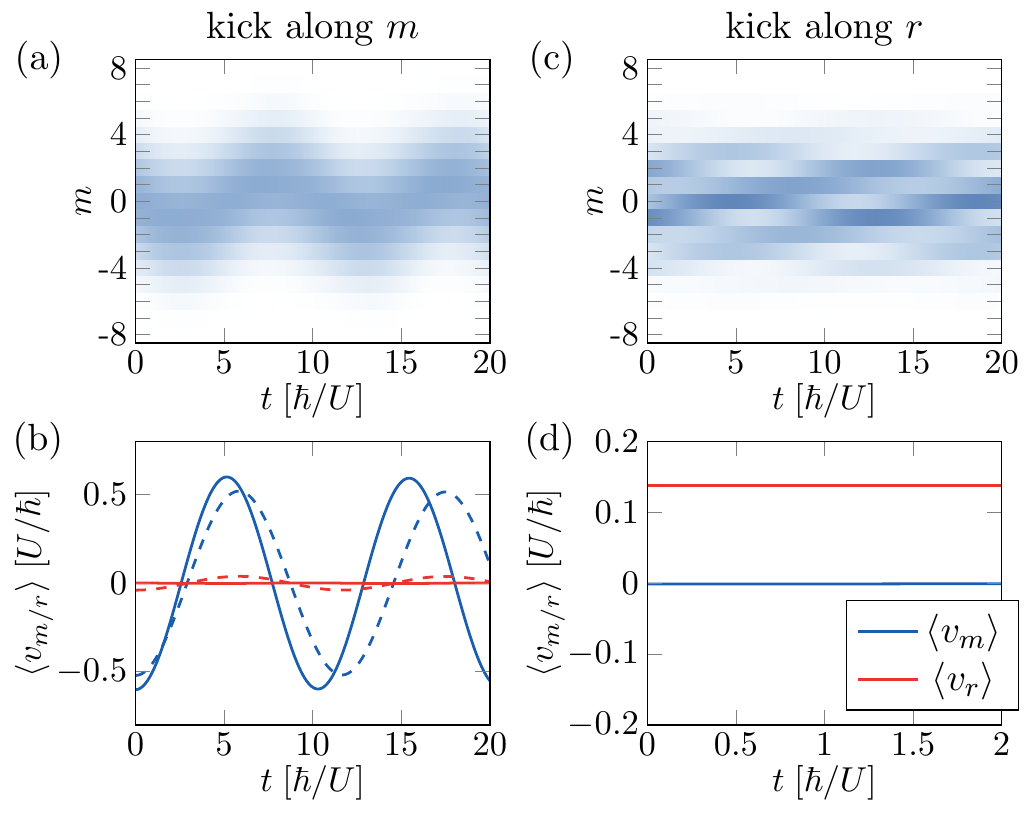}
 \caption{
 Simulated spin dynamics following a velocity kick along $m$ or $r$ for $U_b/U_a=0.5$ (left and right columns, respectively). (a), (c) Evolution of the spin projection probabilities $\Pi_m$. (b), (d) Evolution of the mean velocities $\langle v_m\rangle$ (blue lines) and $\langle v_r\rangle$ (red lines) for the same evolutions. The dashed lines in (b) are calculated with $U_b/U_a=0.4$.
 \label{fig_dynamics}}
 \end{figure}

\subsection{Decoupling of  $m$ and $r$ dynamics}

We show in \fig{fig_dynamics} the dynamics subsequent to the $m$ and $r$ velocity kicks, for $U_b=U_a/2$. 

For an excitation along $m$, the projection probabilities $\Pi_m$ and the mean velocity  $\langle v_m\rangle$ oscillate  consistently with harmonic trapping along $m$. The oscillation frequency matches the value of $\omega$ given in \eqP{eq_omega}. In contrast, the mean velocity $\langle v_r\rangle$  remains close to zero. 

An opposite behavior occurs for an excitation along $r$: the spin distribution $\Pi_m$ becomes modulated in $m$ with a period 3, and this modulation coherently evolves in time with a given chirality. The mean velocity $\langle v_r\rangle$ remains stationary at a non-zero value, consistently with the absence of trapping along $r$. The mean velocity $\langle v_m\rangle$ remains close to zero. These two evolutions are thus consistent with independent dynamics of the two coordinates $m$ and $r$.

We also studied the effect of a departure from the relation  $U_b/U_a=0.5$ by repeating the simulation with $U_b/U_a=0.4$ (dashed lines in \fig{fig_dynamics}(b)).  For an excitation along $m$, we obtain a non-zero oscillation of $\langle v_r\rangle$, which confirms  that the $m$ and $r$ dynamics are rigorously decoupled  under the condition $U_b/U_a=0.5$ only, as found in section \ref{section_protocol}. Nevertheless, we expect that the interpretation of spin dynamics in terms of motion in two dimensions remains valid away from the condition $U_b/U_a=0.5$, albeit with $m$ and $r$ not orthogonal. 

\section{Implementation with cold atoms}

\subsection{Implementation with lanthanide atoms}
This proposal requires using an atomic species with an internal spin $J\gg1$. Lanthanide atoms exhibit a large electronic spin in the ground state, namely $J=8$, $J=6$ and $F=4$ for dysprosium, erbium and thulium -- the species brought to quantum degeneracy so far \cite{lu_strongly_2011,aikawa_bose-einstein_2012,davletov_machine_2020}. The levels spectra shown in \fig{fig_spectrum} and the low-energy dynamics shown in \fig{fig_dynamics} were calculated for $J=8$, and are thus relevant for a practical implementation with dysprosium atoms. Fermionic isotopes of erbium and dysprosium, which were also produced in the quantum degeneracy regime \cite{lu_quantum_2012,aikawa_reaching_2014}, feature a hyperfine structure with an even larger total spin length.

The spin couplings involved in the Hamiltonian \eqP{eq_H} can be implemented using the AC-Stark shift produced by off-resonant lasers \cite{grimm_optical_2000}. In general, second-order light shifts produce spin couplings described by tensors of rank 0, 1 and 2 \cite{cohen-tannoudji_experimental_1972}. For alkali or two-electron atoms, the electronic ground state is isotropic ($s$ valence shell with an orbital angular momentum $L=0$), prohibiting spin-dependent light shifts. Spin transitions can arise from higher-order processes involving the fine or hyperfine couplings, albeit with significant values only close to optical resonances  \cite{mathur_light_1968}. Lanthanide atoms exhibit a more favorable electronic structure for the realization of spin-dependent light shifts, thanks to the  anisotropic electronic orbitals in their electronic ground state. The interaction with light inherits  a significant spin dependency from this anisotropy, even for light far detuned from resonances \cite{lepers_anisotropic_2014}. Furthermore, spin couplings can be further enhanced using light close to a single narrow optical transition \cite{kao_anisotropic_2017}.

In practice, the spin couplings can be produced using resonant optical transitions in the presence of a quantization magnetic field along $z$. Denoting $\omega_{\text L}$ the Larmor frequency, a two-photon process involving two light frequencies of difference $\Delta\omega$ will produce a first- (second-) order spin coupling for $\Delta\omega=\omega_{\text L}$ ($\Delta\omega=2\omega_{\text L}$, respectively). An important asset of this protocol is its protection from magnetic field fluctuations. Indeed, the $\ket{r,n}$ basis states are not magnetized along $z$ (see \fig{fig_spectrum}c), such that magnetic field perturbations cancel at first order.

\subsection{Coupling to a spatial dimension: \\example of a quantum Hall cylinder\label{section_xr_cylinder}}
When the spin couplings are induced by two-photon optical transitions from a single laser spatial mode, they are not coupled to the atom motion. The dynamics can be enriched when they involve light beams propagating along different directions, such that spin transitions occur together with a momentum kick exchanged with light. We present in this section an application of such a spin-orbit coupling, yielding dynamics mimicking a quantum Hall cylinder, with an additional harmonic degree of freedom. We mention that quantum Hall cylinders have been recently realized by directly coupling a small number of spin levels \cite{li2018bose,han_band_2019,liang_coherence_2021}. 

We assume the spin couplings to be driven by two-photon optical transitions using a pair of  laser beams counter-propagating along the spatial coordinate $x$. The couplings then inherit the complex phase factor $\E^{2\I kx}$ from the laser beam interference, where $k$ is the light momentum. The atom dynamics is governed by the Hamiltonian
\begin{align}
H&=\frac{p_x^2}{2M}+V,\label{eq_Hx}\\
V&=-\left[U_a\frac{J_+}{J}+U_b\frac{J_-^2}{J(J-1/2)}\right]\E^{-2\I kx}+\hc,
\end{align}
where $p_x$ is the $x$-momentum and $M$ is the atom mass.
The two processes increasing the remainder $r$ thus acquire a common phase factor $\E^{-2\I kx}$, leading to a gauge field in the $xr$ plane. On the contrary, the two processes increasing the projection $m$ have opposite phase factors $\E^{\pm2\I kx}$, with a zero mean effect for $U_a=2\,U_b$. Under this condition that we assume in the following, we do not expect the occurrence of an effective magnetic field in the $xm$ plane. Therefore, we expect the system to behave as a quantum Hall cylinder in the two variables $(x,r)$, with another degree of freedom $m$ acting as the coordinate of an independent harmonic oscillator (from the term $n\hbar\omega$ in the spectrum \eqP{eq_spectrum_phi_n}).

\begin{figure}[!t]
 \includegraphics[
  trim={2mm 2mm 0 0.cm},scale=0.95
 ]{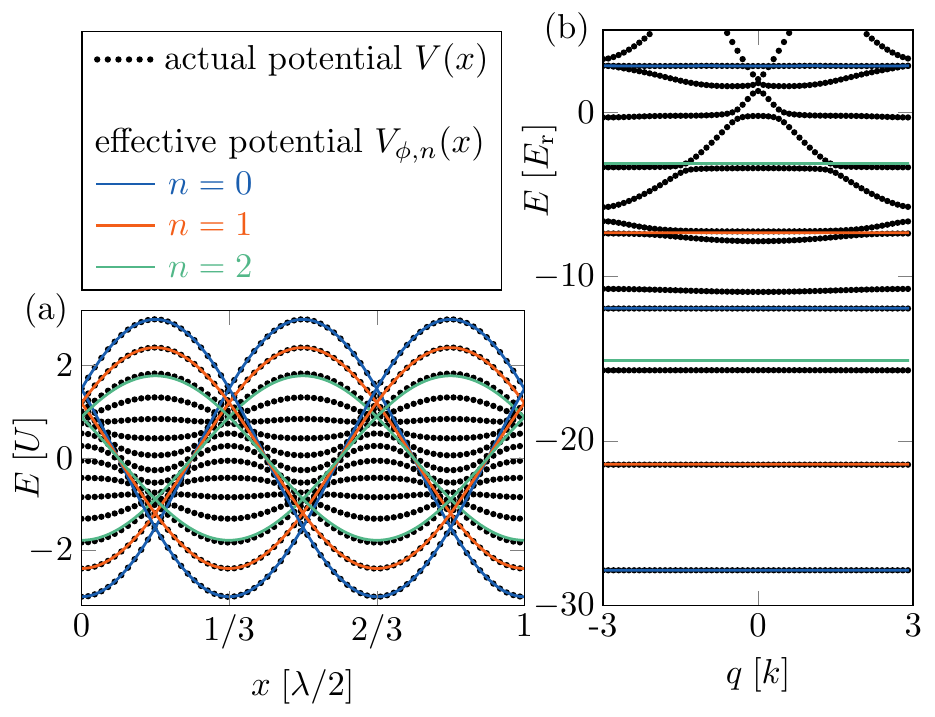}
 \caption{
 (a) Energy spectrum of the coupling $V(x)$ as a function of $x$ (black dots), compared with the effective potentials $V_{\phi,n}(x)$ with $n=0,1,2$ (blue, red and green lines). (b) Bandstructures calculated using the actual potential $V(x)$ (black dots) and the effective ones $V_{\phi,n}(x)$, with $n=0,1,2$ (blue, red and green lines). 
\label{fig_xr_cylinder}}
\end{figure}

In order to reveal this behavior, we generalize the semi-classical treatment discussed above. For each position $x$, we calculate the semi-classical energy functional
\begin{equation}
 V_{\text{cl}}(\theta,\phi,x)=-2U\sin\theta\cos(\phi-2kx)-U\sin^2\theta\cos(2\phi+2kx),
\end{equation}
which always features three extrema for the same orientations, namely $\theta=\pi/2$ and $\phi=0$, $2\pi/3$ or $4\pi/3$.
Expanding the spin operators around these three extrema, we obtain the harmonic spectra
\begin{equation}
 V_{\phi,n}(x)=(E_0+n\hbar\omega )\cos(\phi-2kx),\label{eq_V_phi_n}
\end{equation}
which we compare to the $x$-dependent eigenstates of $V(x)$ in \fig{fig_xr_cylinder}(a). We find an excellent agreement for $n=0$ and 1, and observe a visible departure for $n=2$, signaling the onset of anharmonic effects. 

The energies $V_{\phi,n}(x)$ plays the role of cosine lattice potentials, the angle $\phi$ defining the $x$-position of the energy minima.  Importantly, the three angles $\phi$ play a symmetric role, such that they are all involved in the effective low-energy dynamics -- contrary to the purely spin dynamics studied in section \ref{section_spectrum}. 

The dynamics induced by the potentials  $V_{\phi,n}(x)$ on the $r$ coordinate is better visualized in  the $\ket{r,n}$ position state basis, as
 \begin{align}
 \Veff&=\sum_{n\geq 0}\sum_{\phi=0,\tfrac{2\pi}{3},\tfrac{4\pi}{3}} (E_0+n\hbar\omega)\cos(\phi-2kx)\ket{\phi,n}\bra{\phi,n}\nonumber\\
 &\!\!=\sum_{n\geq 0}\sum_{r=0}^2\frac{E_0+n\hbar\omega }{2}\E^{-2\I kx}\ket{r+1,n}\bra{r,n}+\hc.\label{eq_Vx}
\end{align}
This potential describes hopping dynamics along $r$, with an $x$-dependent complex phase that mimics the Aharonov-Bohm phase associated to a magnetic field in the $xr$ plane. The full atom dynamics, described by the effective Hamiltonian $H_{\text{eff}}=p_x^2/2M+\Veff$, then maps to the motion of a charged particle on a Hall cylinder along $x$ and $r$, with  an additional harmonic degree of freedom $n$.

We validate this description by comparing the energy level structure of the actual Hamiltonian \eqP{eq_Hx} with the effective model \eqP{eq_Vx}. Both models are invariant upon the discrete magnetic translation 
\begin{equation}
 T_{\text{mag}}=T_{x,\lambda/6}R_{z,-2\pi/3},
\end{equation}
which combines a $\lambda/6$-translation along $x$ and rotation of the spin  around $z$ of angle $-2\pi/3$. This symmetry leads to the conservation of the quasi-momentum
\begin{equation}
 q\equiv \frac{M v_x}{\hbar}+2k J_z\Mod{6k},
\end{equation}
defined over the magnetic Brillouin zone $-3k\leq q<3k$. The Hamiltonian spectra organize in magnetic Bloch bands, shown in \fig{fig_xr_cylinder}(b) for a coupling strength $U=12\,\Er$, where $\Er=\hbar^2k^2/(2M)$ is the single-photon recoil energy. The spectrum of the Hamiltonian \eqP{eq_Hx} exhibits very flat lowest energy bands, well reproduced by the bands of the effective model for $n=0,1,2$. This comparison confirms the relevance of the description of low-energy dynamics as the one of a quantum Hall cylinder.

\section{Conclusion}
To conclude, we have shown that, by combining first- and second-order spin couplings, one can simulate two-dimensional dynamics within a large-size atomic spin. This technique extends the synthetic dimension toolbox, and could be applied to simulate various types of topological systems. We described the extension of the method to engineer a quantum Hall cylinder with an additional harmonic degree of freedom. The simulation of two-dimensional dynamics in a single spin will become even more useful for  realizing other types of topological systems in higher dimensions $D>3$, such as four-dimensional quantum Hall systems \cite{price_four-dimensional_2015} or five-dimensional Weyl semi-metals \cite{lian_five-dimensional_2016}. Our method could also be applied to other physical platforms making use of synthetic dimensions \cite{ozawa_topological_2019}.

\section*{Acknowledgments}
We thank Jean Dalibard for stimulating discussions and careful reading of the manuscript.  This work is supported by  European Union (grant TOPODY 756722 from the European Research Council). 

\appendix*

\section*{Appendix: low-energy dynamics}

We give an alternative derivation of the low-energy dynamics of the Hamiltonian \eqP{eq_H}. The ground state obtained from the semi-classical analysis is the coherent spin state polarized along $x$. We use a Holstein-Primakoff transform to express the spin operators in terms of a bosonic degree of freedom \cite{holstein_field_1940}, as 
\begin{align}
 J_x&=J-a^\dagger a,\\
 J_z-\I J_y&=\sqrt{2J-a^\dagger a}\,a,\\
 J_z+\I J_y&=a^\dagger\sqrt{2J-a^\dagger a},
\end{align}
where $a$ is a bosonic annihilation operator. To lowest order, the $z$ spin component
\begin{equation}
J_z\simeq\sqrt{\frac{J}{2}}(a+a^\dagger)
\end{equation}
maps to the position operator of the harmonic oscillator associated with $a$.
Expanding the Hamiltonian in power series in $1/J$, we obtain at first order
\begin{equation}\label{eq_H_quadratic}
 H/U\simeq -3+\frac{10a^\dagger a-a^2-a^{\dagger2}}{2J}.
\end{equation}
This quadratic Hamiltonian can be diagonalized using a Bogoliubov transform, by defining new bosonic operators
\begin{align}
 b&=u\,a+v\,a^\dagger,\\
 b^\dagger&=v^*a+u^*a^\dagger,
\end{align}
with $u^2-v^2=1$. For $u=(1/2+5/(4\sqrt6))^{1/2}\simeq1.005$ and $v=-\sqrt{u^2-1}\simeq-0.102$, the Hamiltonian takes the canonical form
\begin{equation}
H=E_0+\hbar\omega\,b^\dagger b,
\end{equation}
with 
\begin{align}
E_0&=\left(-3+\frac{2\sqrt6-5}{2J}\right)U,\\
\hbar\omega&=\frac{2\sqrt6}{J}U.
\end{align}
This expansion can be reproduced around the semi-classical energy maxima, leading to the complete harmonic spectrum \eqP{eq_spectrum_phi_n} discussed in the main text.

\footnotetext[2]{We consider here the case of an integer spin length $J$, but the proposal can be straightfowardly extended to a half-integer spin.}


%

\end{document}